\documentclass[conference]{IEEEtran}
\IEEEoverridecommandlockouts
\usepackage{cite}
\usepackage{amsmath,amssymb,amsfonts}
\usepackage{algorithmic}
\usepackage{algorithm}
\usepackage{graphicx}
\usepackage{textcomp}
\usepackage{xcolor}
\usepackage{booktabs}
\usepackage{setspace}
\usepackage{multirow}
\usepackage{svg}
\graphicspath{{69ba2833d399e6c1053a9882/}}
\usepackage{subcaption}
\usepackage{tikz}
\usepackage{pgfplots}
\usepackage[hidelinks]{hyperref}
\pgfplotsset{compat=1.18}
\usepgfplotslibrary{groupplots}
\usetikzlibrary{arrows.meta,backgrounds,calc,fit,matrix,positioning}
\def\BibTeX{{\rm B\kern-.05em{\sc i\kern-.025em b}\kern-.08em
    T\kern-.1667em\lower.7ex\hbox{E}\kern-.125emX}}
\begin{document}
\pagestyle{plain}

\title{FusionRCG: Orchestrating Recursive Computation Graphs across GPU Memory Hierarchies%
\thanks{\IEEEauthorrefmark{1}These authors contributed equally to this work.}
\thanks{
Corresponding authors: Fusong Ju (jufusong@bza.edu.cn), 
Wei Hu (whuustc@ustc.edu.cn),
Huanhuan Xia (huanhuanxia@zgci.ac.cn).
}
}

\author{
\IEEEauthorblockN{
Yihong Zhang\IEEEauthorrefmark{1}\IEEEauthorrefmark{4}\IEEEauthorrefmark{2},
Xinran Wei\IEEEauthorrefmark{1}\IEEEauthorrefmark{2}\IEEEauthorrefmark{5},
Junshi Chen\IEEEauthorrefmark{4},
Fusong Ju\IEEEauthorrefmark{2},
Wei Hu\IEEEauthorrefmark{4},
Jinlong Yang\IEEEauthorrefmark{4},
Huanhuan Xia\IEEEauthorrefmark{2}\IEEEauthorrefmark{5}
}

\IEEEauthorblockA{\IEEEauthorrefmark{4}\textit{University of Science and Technology of China}, Hefei, Anhui, China}
\IEEEauthorblockA{\IEEEauthorrefmark{2}\textit{Zhongguancun Academy}, Beijing, China}
\IEEEauthorblockA{\IEEEauthorrefmark{5}\textit{Zhongguancun Institute of Artificial Intelligence}, Beijing, China}
}
\maketitle
\bstctlcite{BSTcontrol}
\setstretch{1.055}

\begin{abstract} 
Evaluating high-dimensional integrals via deep hierarchical recurrences is a dominant cost in quantum chemistry. While CPUs manage these efficiently, GPUs suffer a critical mismatch: limited per-thread memory is quickly overwhelmed by an explosion of simultaneously live intermediate variables. As recurrence scales, this forces massive data spilling to global memory, collapsing performance into a severe memory-bound regime. We present FusionRCG, a framework that jointly optimizes computation graph structure and GPU memory mapping. Exploiting the inherent topological flexibility of recurrence graphs, using electron repulsion integrals as an example, we contribute: (1) liveness-aware graph orchestration to minimize peak live intermediates; (2) algebraic dimensionality reduction via stepwise Cartesian-to-spherical fusion, shrinking intermediate footprints by up to $7.7\times$; and (3) an adaptive multi-tier kernel architecture routing graphs across the memory hierarchy. Evaluated on NVIDIA A100 GPUs, FusionRCG achieves up to $3.09\times$ end-to-end SCF speedup over GPU4PySCF and maintains $75\%$ parallel efficiency at 64~GPUs, successfully rescuing these workloads from memory-bound limits.
\end{abstract}


\section{Introduction}

In modern large-scale quantum chemistry simulations, evaluating high-dimensional tensor integrals via deep hierarchical recurrences constitutes a critical computational bottleneck~\cite{b2,b4,b23}. From an algorithmic and computer architecture perspective, such workloads can be naturally modeled as \textbf{\emph{Recursive Computation Graphs}} (RCGs), where vertices represent intermediate quantities and edges encode recurrence dependencies. As the polynomial degree or the dimensionality of the problem increases, the number of intermediate variables that must be simultaneously maintained within this computation graph undergoes a combinatorial explosion. While CPUs can efficiently manage this super-polynomial growth by utilizing sophisticated out-of-order execution engines and deep, multi-level cache hierarchies to absorb the exploding intermediate state~\cite{b10,b8}, porting such recursive workloads to modern GPUs exposes a severe architectural mismatch. GPU performance relies on massive thread-level parallelism, which strictly constrains per-thread on-chip storage~\cite{b17}. When the working set of simultaneously live intermediates breaches the hard ceiling of physical registers, the compiler is forced to spill massive amounts of data to the high-latency, off-chip global memory~\cite{b7}. This pure memory-hierarchy crisis drastically collapses hardware thread occupancy, submerging the execution in memory access latency and degrading an inherently compute-bound kernel into a memory-bound bottleneck~\cite{b5}.

\begin{figure}[t]
\centering
\includegraphics[width=\columnwidth]{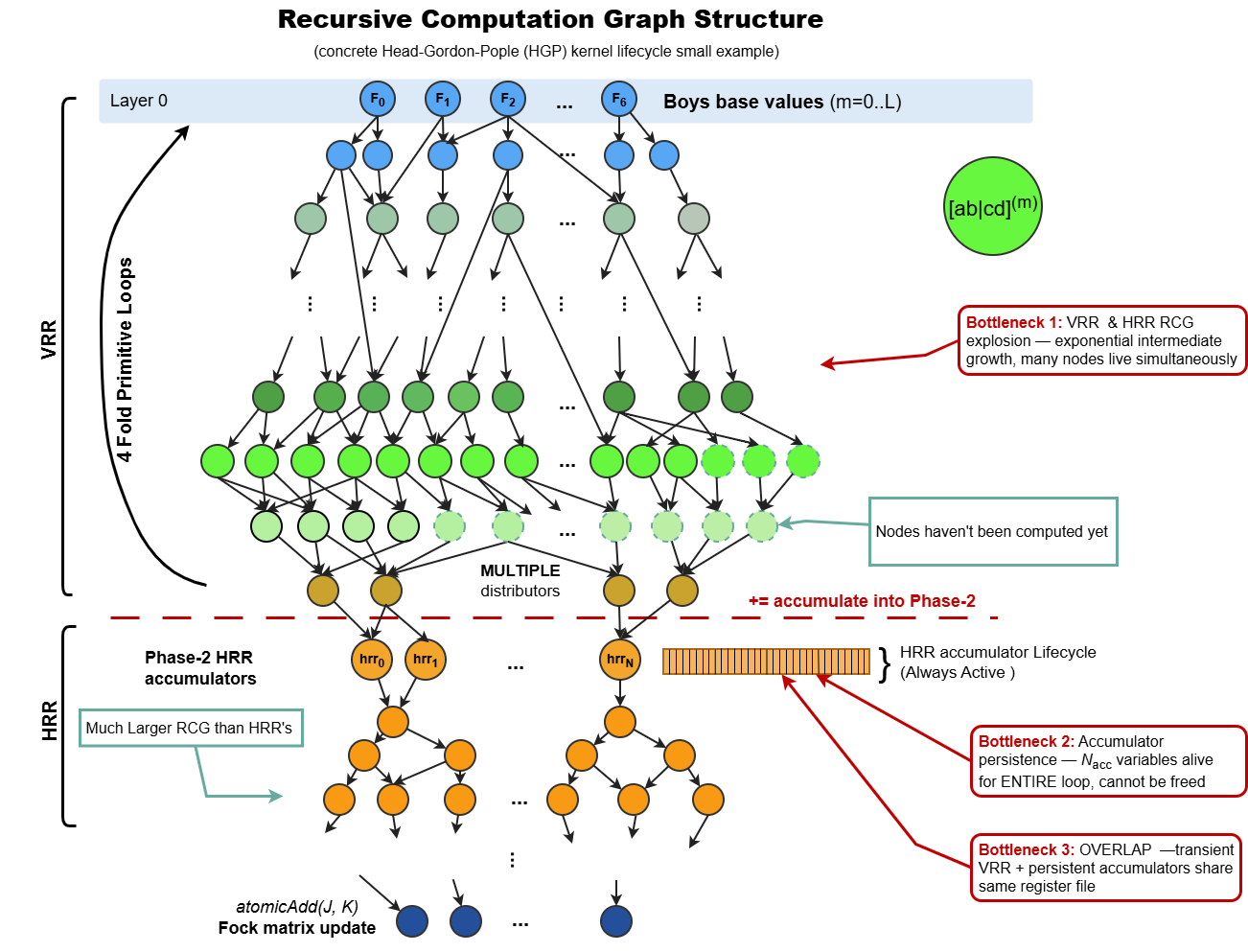}
\caption{Recursive computation graph structure of HGP and its associated GPU bottlenecks. The first recurrence phase generates a rapidly expanding dependency graph with a large peak-live frontier, while the second recurrence phase introduces long-lived accumulators; these phases are formalized later as VRR and HRR, respectively, in Section~\ref{sec:hgp}. Their overlap within the same register file creates the fundamental memory bottleneck.}
\label{fig:hgp-bottleneck}
\end{figure}

A representative and highly demanding instantiation of this memory-bound RCG is the calculation of electron repulsion integrals (ERIs). ERIs are 4-index tensor integrals that describe particle interactions, serving as a foundational operator in many high-fidelity scientific simulations and typically consuming 80\% of total runtime~\cite{b2,b3,b4,b23}. Accelerating ERI calculation has therefore been a long-standing objective within the computational chemistry community, and existing GPU engines have achieved notable success~\cite{b3,b4}. Most successful GPU implementations are based on Rys quadrature or McMurchie--Davidson (MD) formulations~\cite{b3,b24,b25}. These formulations are well aligned with SIMT execution because they expose regular batched parallelism and maintain a comparatively controlled intermediate footprint.

The Head-Gordon-Pople (HGP) algorithm, by contrast, represents a markedly different ERI formulation. HGP and its variants remain highly influential in CPU engines due to their favorable contraction-aware arithmetic structure and their substantially lower floating-point operation count than Rys quadrature and McMurchie--Davidson formulations~\cite{b2,b26,b27}. However, executing HGP efficiently on GPUs is notoriously difficult, and mainstream adoption has been severely limited. This difficulty is fundamentally driven by HGP's mathematical formulation: it evaluates an ERI through deep, chained recurrences over Cartesian intermediates. When mapped onto SIMT hardware, a target integral expands into a massive dependency graph where many intermediates are produced early but consumed much later. This forces an enormous working set to remain simultaneously live, triggering the exact memory-hierarchy crisis described above. Because of this severe register pressure, massive data spilling, and excessive global-memory traffic, many optimization efforts for HGP on GPUs inevitably hit a performance wall. Figure~\ref{fig:hgp-bottleneck} illustrates this recursive computation graph structure and the memory bottlenecks during two HGP phases (formalized in Section~\ref{sec:hgp}). 

Bridging the gap between HGP's arithmetic advantages and its limited practical viability on GPUs requires more than accelerating a fixed kernel. The key insight of this work is that a strictly defined mathematical result does not imply a unique computation graph. This reveals a previously under-exploited optimization space in which HGP can be analyzed along three tightly coupled dimensions relevant to GPU execution:

\medskip
\noindent\quad$\bullet$~\textbf{Computation Graph Structure and Execution Order.} The HGP recurrence does not induce a unique computation graph. Different legal reduction paths generate structurally distinct graphs, and different topological orders expose different live ranges even for the same graph. Together, these factors determine the peak number of simultaneously live intermediates.

\medskip
\noindent\quad$\bullet$~\textbf{Intermediate State Size Optimization.} The cost of HGP on GPUs is governed not only by arithmetic count, but also by the physical size of the intermediate state carried between recurrence stages and contraction. The basis and representation in which these intermediates are maintained directly affect both storage demand and data movement.

\medskip
\noindent\quad$\bullet$~\textbf{Memory Mapping across GPU Hierarchy.} As angular momentum increases, the working set of the recurrence inevitably outgrows the smallest on-chip storage. Performance therefore depends on how intermediate state is placed across registers, shared memory, and global memory, rather than on arithmetic optimization alone.

Based on these insights, we present \textbf{FusionRCG}, a framework that addresses the memory wall for hierarchical recursive computation graphs by systematically orchestrating their evaluation. By co-designing the graph structure, memory mapping, and algebraic boundaries, FusionRCG preserves the exact mathematical arithmetic of HGP while eliminating its historical GPU bottlenecks. 

Our contributions are threefold, directly addressing the dimensions above:

\medskip
\noindent\quad$\bullet$~\textbf{Liveness-aware graph orchestration for register-contained execution} (\S\ref{sec:orchestration}). By jointly optimizing the graph topology and the evaluation schedule, FusionRCG drastically minimizes the peak working set of live intermediates. This structural optimization effectively eliminates register spilling for low-to-moderate degree recurrences, keeping the dense execution entirely on-chip.

\medskip
\noindent\quad$\bullet$~\textbf{Algebraic dimensionality reduction to minimize global-memory traffic} (\S\ref{sec:spherical}). We step-wise fuse the standard Cartesian-to-pure-spherical shell transformation~\cite{b8,b9} directly into the final stages of the recurrence. This shrinks the contraction-bound intermediate tensors and the associated buffer writes / atomicAdd traffic by up to $7.7\times$ without introducing any mathematical approximation.

\medskip
\noindent\quad$\bullet$~\textbf{An adaptive multi-tier execution strategy across the GPU memory hierarchy} (\S\ref{sec:tiered}). We demonstrate that memory pressure arises from distinct sources as the problem scales. To address this, the framework automatically routes each computation graph to the appropriate memory level---emitting fundamentally different kernel architectures (register-only or shared-memory-buffered) based on computational complexity.

\begin{figure*}[t]
\includegraphics[width=0.95\textwidth]{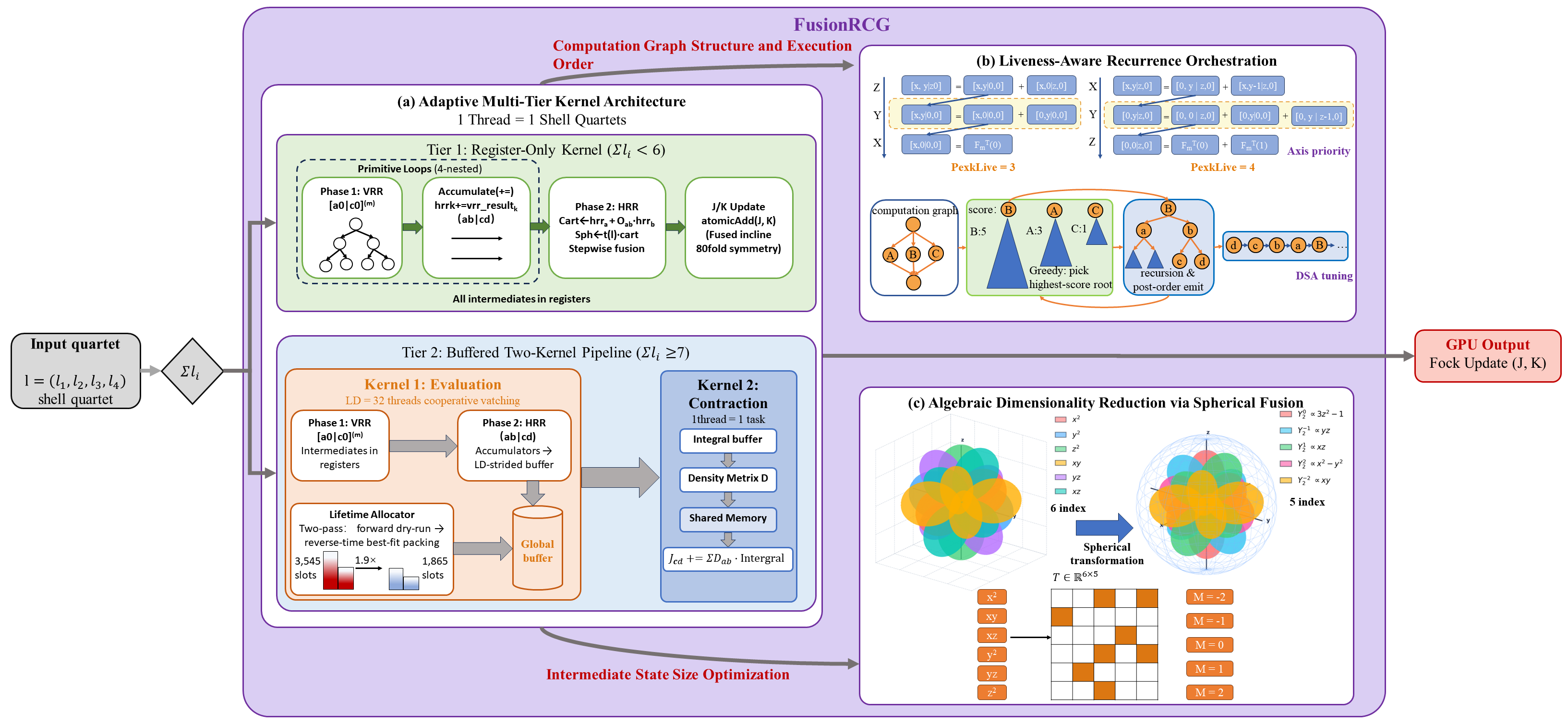}
\caption{FusionRCG overview. Given an angular-momentum quartet, the generator first optimizes the Phase~1 VRR graph for low peak liveness, then selects a register-only Tier~1 backend or a buffered Tier~2 backend based on decomposed register demand, and finally fuses the standard Cartesian-to-pure-spherical shell transformation into the Phase~2 HRR stage before generating specialized CUDA kernels.}
\label{fig:overview}
\end{figure*}

Evaluated on NVIDIA A100 GPUs against the state-of-the-art GPU4PySCF~\cite{b3}, FusionRCG achieves up to $3.09\times$ end-to-end SCF speedup on representative molecular systems with an average of $2.4\times$ at the cc-pVQZ level, and scales to 64~GPUs with $75\%$ parallel efficiency, successfully rescuing the execution from the memory-bound regime. Figure~\ref{fig:overview} provides an overview of the code generation and memory mapping pipeline.

\section{Background and Motivation}
\label{sec:background}

\subsection{The HGP Recurrence for ERIs}
\label{sec:hgp}

To understand the origin of the memory bottleneck, we first examine the underlying mathematical formulation. An electron repulsion integral over four Gaussian basis functions $\mathbf{a},\mathbf{b},\mathbf{c},\mathbf{d}$ is defined as:
\begin{equation}
(\mathbf{ab}|\mathbf{cd}) = \iint \frac{\phi_\mathbf{a}(\mathbf{r}_1)\phi_\mathbf{b}(\mathbf{r}_1)\,\phi_\mathbf{c}(\mathbf{r}_2)\phi_\mathbf{d}(\mathbf{r}_2)}{|\mathbf{r}_1 - \mathbf{r}_2|}\,d\mathbf{r}_1\,d\mathbf{r}_2
\label{eq:eri-def}
\end{equation}
where $\phi_\mathbf{a}$ is a Cartesian Gaussian with angular momentum $\mathbf{a} = (a_x, a_y, a_z)$, and total angular momentum $|\mathbf{a}| = l_a$. In the notation below, $(\mathbf{ab}|\mathbf{cd})$ denotes the target contracted ERI, whereas the bracketed quantities $[\cdots|\cdots]^{(m)}$ denote the HGP auxiliary integral family indexed by the Boys order $m$. The HGP algorithm~\cite{b2} evaluates Eq.~\eqref{eq:eri-def} through a two-phase hierarchical recurrence:

\medskip
\noindent\textbf{Phase~1: Vertical Recurrence Relation (VRR).}
Starting from Boys-function base values $F_m(T) \equiv [\mathbf{0}\,\mathbf{0}|\mathbf{0}\,\mathbf{0}]^{(m)}$, VRR builds up angular momentum one index component at a time:
\begin{align}
[\mathbf{a}+\mathbf{1}_i\;\mathbf{0}|\mathbf{c}\;\mathbf{0}]^{(m)} &= (P_i - A_i)\,[\mathbf{a}\;\mathbf{0}|\mathbf{c}\;\mathbf{0}]^{(m)} \nonumber \\
&\quad + (W_i - P_i)\,[\mathbf{a}\;\mathbf{0}|\mathbf{c}\;\mathbf{0}]^{(m+1)} \nonumber \\
&\quad + \frac{a_i}{2\zeta}\bigl([\mathbf{a}-\mathbf{1}_i\;\mathbf{0}|\mathbf{c}\;\mathbf{0}]^{(m)} - \frac{\rho}{\zeta}[\cdots]^{(m+1)}\bigr) \nonumber \\
&\quad + \frac{c_i}{2(\zeta+\eta)}\,[\mathbf{a}\;\mathbf{0}|\mathbf{c}-\mathbf{1}_i\;\mathbf{0}]^{(m+1)}
\label{eq:vrr}
\end{align}
where $\mathbf{1}_i$ is the unit vector along axis $i$, and $\zeta, \eta, \rho, \mathbf{P}, \mathbf{W}$ are Gaussian-product parameters. Each node depends on 2 to 5 parents. Crucially, the choice of \emph{which} index $i$ to increment is not strictly fixed, providing a structural degree of freedom that we will exploit.

\medskip
\noindent\textbf{Phase~2: Horizontal Recurrence Relation (HRR).}
After contraction, HRR transfers angular momentum from center-pair products to individual centers via a simpler binary recurrence:
\begin{equation}
(\mathbf{a}\;\mathbf{b}+\mathbf{1}_i|\mathbf{c}\;\mathbf{d}) = (\mathbf{a}+\mathbf{1}_i\;\mathbf{b}|\mathbf{c}\;\mathbf{d}) + (A_i - B_i)\,(\mathbf{a}\;\mathbf{b}|\mathbf{c}\;\mathbf{d})
\label{eq:hrr}
\end{equation}
Each HRR node depends on exactly 2 parents. While the operation count of this two-phase approach is highly efficient, executing its data dependencies on massively parallel hardware exposes severe physical constraints.

\subsection{GPU Memory Hierarchy and the Register Wall}
\label{sec:gpu-mem}

Modern GPUs derive their massive throughput from the Single Instruction Multiple Threads (SIMT) execution model, which requires maintaining the architectural state of thousands of in-flight threads simultaneously. To support this, NVIDIA GPUs provide a massive but strictly partitioned register file. Each Streaming Multiprocessor (SM) contains 65,536 32-bit registers shared among all resident threads, imposing a hard per-thread architectural ceiling of 255 registers.

When a compute kernel's working set exceeds this 255-register limit, the compiler is forced to spill excess intermediate values to per-thread \emph{local memory}, which is physically backed by off-chip High Bandwidth Memory (HBM). These spills are devastating to performance because the bandwidth gap across the memory hierarchy is extreme. The SM register file delivers an aggregate bandwidth of ${\sim}19$\,TB/s, whereas global HBM provides only ${\sim}2$\,TB/s---a disparity of nearly $10{,}000\times$ at the per-SM level. Between these two extremes, the shared memory/L1 cache ($19$\,TB/s, $164$\,KB/SM) and L2 cache ($5$\,TB/s, $40$\,MB) offer vital intermediate capacity-bandwidth trade-offs. FusionRCG utilizes this exact hierarchy, mapping intermediates strictly to registers (Tier~1) or buffering them through shared memory and global memory (Tier~2) based on the computation's exact spatial footprint.

\subsection{From Recurrence to Register Wall}
\label{sec:dag-analysis}

When the HGP mathematical formulation is mapped onto the SIMT execution model, it manifests as a dense computation graph whose size grows combinatorially with the angular-momentum quartet $\mathbf{l} = (l_a, l_b, l_c, l_d)$. Because each node in this graph represents a double-precision intermediate occupying two 32-bit registers, even modest quantum systems generate massive register demand. 

For example, at $\mathbf{l}=(1,1,1,1)$ ($\sum l_i = 4$), the graph contains over 700 nodes. If all intermediates were simultaneously live, the naive register requirement would exceed the 255-register hardware limit by nearly $6\times$. By $\mathbf{l}=(2,2,2,2)$ ($\sum l_i = 8$), the graph grows to thousands of nodes, surpassing the limit by more than $28\times$. 

\begin{figure}[t]
\centering
\includegraphics[width=\columnwidth]{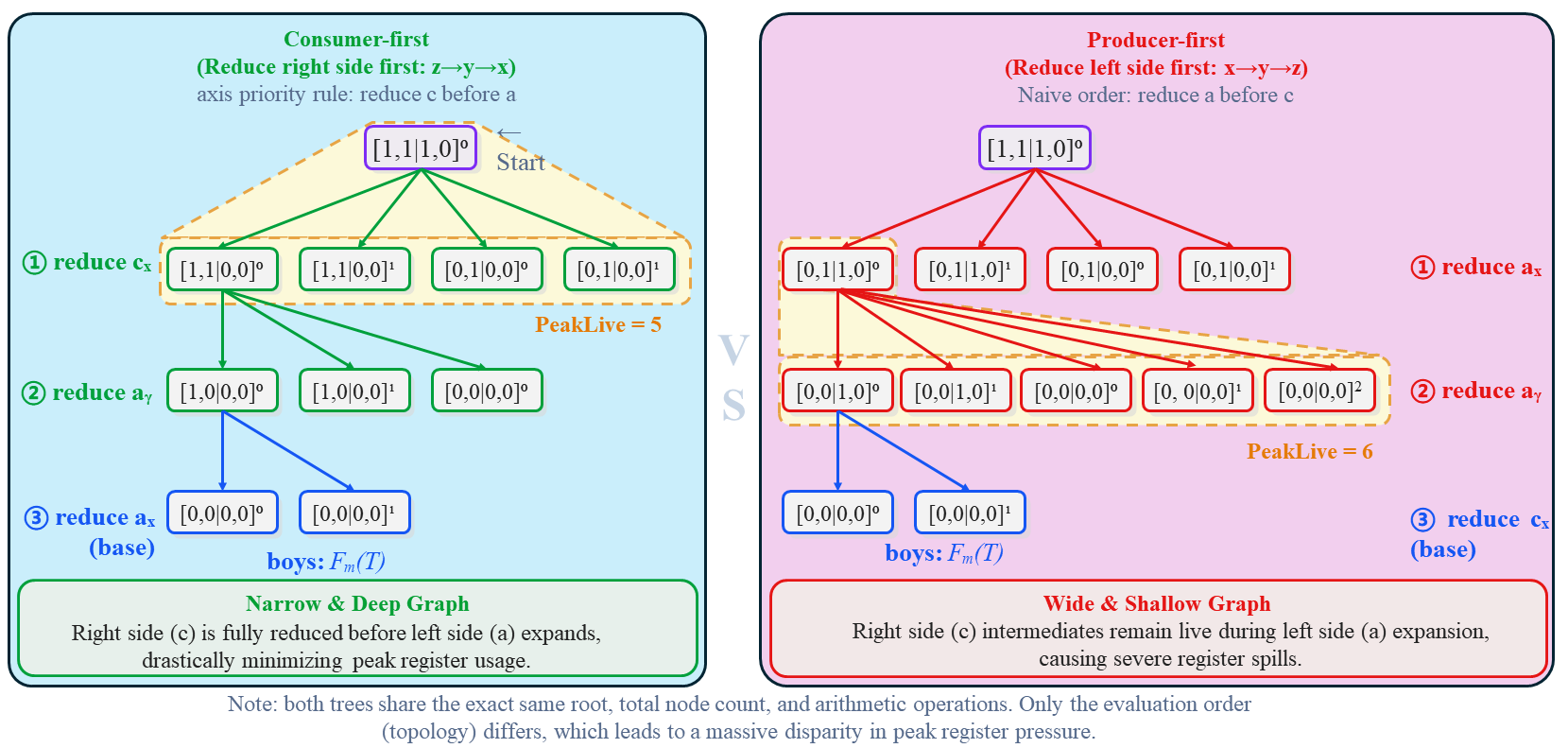}
\caption{Axis selection reshapes the VRR computation graph \emph{before} scheduling. Starting from the same recurrence state, consumer-first (left) and producer-first (right) generate graphs with identical arithmetic but different frontier width. The consumer-first graph is narrower, so fewer intermediates must remain live simultaneously.}
\label{fig:axis}
\end{figure}

Fortunately, the hardware does not need to store the entire graph simultaneously. A register is occupied only while the corresponding intermediate remains needed by a future consumer; once all uses of that value are complete, its register is reclaimed. Therefore, the actual bottleneck within the recurrence graph is the \emph{peak simultaneously live} set. Let $G = (V, E)$ denote the computation graph, where each vertex $v \in V$ is an intermediate and each edge encodes a data dependency. Given a topological ordering $\pi$ of $G$, a vertex $v$ is \emph{live} at step $t$ if it has been computed but at least one of its consumers has not yet been scheduled. The peak liveness is defined as:
\begin{equation}
\mathrm{PeakLive}(\pi) = \max_{t}\;\bigl|\{v \in V \mid v \text{ live at step } t\}\bigr|
\label{eq:peaklive}
\end{equation}
This $\mathrm{PeakLive}$ metric is highly sensitive to both the graph's structural topology and its evaluation order---the two degrees of freedom that FusionRCG optimizes.

However, peak liveness captures only the \emph{transient} portion of the on-chip footprint. In a full HGP kernel, these short-lived VRR intermediates must coexist with a second, long-lived term: the Phase~2 accumulators. These accumulators are allocated before the primitive loops begin and must remain resident across all primitive combinations. For a shell quartet $\mathbf{l}=(l_1,l_2,l_3,l_4)$, their count is:
\begin{equation}
N_\mathrm{acc} = \prod_{i=1}^{4}(2l_i+1)
\label{eq:nacc}
\end{equation}
The total shared register demand is therefore dictated by their superposition:
\begin{equation}
R_\mathrm{total} \;\approx\; 2\bigl(\mathrm{PeakLive}_\mathrm{VRR} + N_\mathrm{acc}\bigr)
\label{eq:rtotal}
\end{equation}
Ultimately, the register wall is not caused by isolated bottlenecks, but by the \emph{overlap} of a massive short-lived VRR frontier colliding with a persistent accumulator bank within the same finite register file. Figure~\ref{fig:hgp-bottleneck} visually summarizes this shared-pressure bottleneck and how the combined footprint dictates the feasible execution regime.

\begin{algorithm}[t]
\caption{Axis-priority rule (consumer-first, $z \to y \to x$).}
\label{alg:axis-priority}
\begin{algorithmic}[1]
\FOR{$i = 0,\, 1,\, 2$}
\IF{$c[i] > 0$}
\STATE $\mathrm{axis} \gets i + 3$ \hfill \textit{// right side first}
\ELSIF{$a[i] > 0$}
\STATE $\mathrm{axis} \gets i$ \hfill \textit{// then left side}
\ENDIF
\ENDFOR
\end{algorithmic}
\end{algorithm}

\section{FusionRCG}
\label{sec:FusionRCG}
The three components of FusionRCG operate at distinct but coupled levels. 
At the mathematical level, HGP exposes flexibility in recurrence path selection and intermediate representation. 
At the algorithmic level, FusionRCG converts this flexibility into graph construction, scheduling, and staged transformation rules during code generation. 
At the architectural level, these decisions reduce register pressure, shrink data movement, and determine whether execution remains register-resident or must be routed through higher memory tiers.

FusionRCG integrates three co-designed optimizations within a single code-generation framework: liveness-aware recurrence orchestration (\S\ref{sec:orchestration}), algebraic dimensionality reduction of the contraction-bound intermediate space (\S\ref{sec:spherical}), and adaptive multi-tier mapping across the GPU memory hierarchy (\S\ref{sec:tiered}).
All decisions in this section are made at code-generation time---the generated GPU kernel contains a fixed, fully unrolled instruction sequence with no runtime branching.

\subsection{Liveness-Aware Recurrence Orchestration}
\label{sec:orchestration}
Mathematically, HGP admits multiple equivalent recurrence paths for the same target integral. Algorithmically, we exploit this freedom through graph construction and topological scheduling at code-generation time. Architecturally, the objective is to minimize the peak live frontier so that Phase~1 remains as register-resident as possible.

The HGP recurrence (Eq.~\ref{eq:vrr}) increments one component of the multi-index $(\mathbf{a}, \mathbf{c})$ per step.
With $\mathbf{a} = (a_x, a_y, a_z)$ and $\mathbf{c} = (c_x, c_y, c_z)$, there are up to six valid reduction axes at each node: three on the left side ($a_x, a_y, a_z$) and three on the right side ($c_x, c_y, c_z$).

Choosing different axes does not change the total node count or the total arithmetic---but it \emph{fundamentally reshapes} the graph topology.
A ``wide'' expansion (reducing high-degree axes first) creates broad wavefronts where many intermediates are simultaneously live.
A ``deep'' expansion (reducing low-order axes first) creates narrow channels that limit concurrent liveness.
Figure~\ref{fig:axis} isolates this \emph{graph-construction} effect: two axis-selection strategies on the same starting node produce computation graphs with identical node count but $1.5$--$2\times$ different PeakLive.
Figure~\ref{fig:overview}(b) then isolates the execution strategy after the computation-graph structure has been fixed.

This observation is the foundation of our approach: \emph{before} optimizing the evaluation order of a fixed graph (a compiler problem), we optimize the \emph{graph structure itself} (a domain-specific algebraic problem).

\medskip
\noindent\textbf{Axis-Priority Rule.}
FusionRCG uses a deterministic axis-priority rule that selects the reduction axis at each VRR node (Algorithm~\ref{alg:axis-priority}).

This ordering is motivated by two observations. First, reducing ket-side indices before bra-side indices (consumer-first) generates subgraphs whose outputs are consumed sooner by the downstream HRR phase, thereby shortening bra-side accumulator lifetimes. Second, within each side, the $z \to y \to x$ traversal order exploits the fact that $z$-components have the smallest fan-out per recursion step in the Cartesian basis ordering, creating narrower wavefronts that minimize simultaneously live nodes.
On $\mathbf{l}=(1,1,1,1)$, this priority reduces PeakLive by 42\% versus the worst axis ordering (bra-first, $x \to y \to z$), with zero change to total arithmetic or node count.

\medskip
\noindent\textbf{Dependency Scheduling Algorithm (DSA).}
Given a fixed computation graph $G = (V, E)$ from the axis-priority construction, we then optimize only the evaluation order on that fixed computation graph. This subproblem is closely related to classical minimum-register sequencing and register-pressure-aware instruction scheduling on trees, computation graphs, and GPU scheduling regions~\cite{b13,b15,b16,b28,b29,b30}. In contrast to Figure~\ref{fig:axis}, which changes the graph structure itself, DSA keeps the nodes and edges unchanged and seeks a topological ordering $\pi$ minimizing $\mathrm{PeakLive}(\pi)$. DSA is therefore not a new scheduling formulation; it is a lightweight domain-specific heuristic tailored to our offline recurrence code generation. Our greedy heuristic maintains a ready set~$\mathcal{R}$ and at each step selects:
\begin{equation}
v^* = \arg\max_{v \in \mathcal{R}}\; \sum_{w \in \mathrm{Reach}(v)} \bigl(\mathrm{out}(w) - \mathrm{in}(w)\bigr)
\label{eq:dsa-score}
\end{equation}

The intuition behind this scoring function is that it captures the ``\emph{register release potential}'' of scheduling~$v$ next.
Nodes reachable from $v$ with high fan-out ($\mathrm{out} \gg \mathrm{in}$) are \emph{distributors}---scheduling them early enables their consumers to execute promptly and release registers.
Nodes with high fan-in ($\mathrm{in} \gg \mathrm{out}$) are \emph{accumulators}---deferring them avoids premature liveness inflation.
DSA thus prioritizes paths that lead to rapid register turnover (Figure~\ref{fig:overview}(b)).

The algorithm runs in $O(|V|^2)$ time, acceptable for offline code generation: the full schedule for $\mathbf{l}=(2,2,2,2)$ ($|V|=3{,}545$) completes in seconds on one CPU core.

\subsection{Algebraic Dimensionality Reduction via Spherical Fusion}
\label{sec:spherical}

Mathematically, the Phase~2 result can be represented either in the Cartesian shell basis or, after an exact shell transformation, in the pure spherical basis. Algorithmically, FusionRCG fuses this basis change into the recurrence boundary instead of materializing the full Cartesian tensor first. Architecturally, this reduces the size of contraction-bound intermediates and the associated global-memory traffic.

Phase~2 naturally produces contraction-bound intermediates resolved in the Cartesian Gaussian shell basis. In most electronic-structure implementations, however, the final shell representation is the pure spherical harmonic Gaussian basis, so an additional Cartesian-to-spherical transformation is required~\cite{b8,b9}.

For a shell of angular momentum $l$, this transformation has the standard form
\begin{equation}
\mathcal{I}_{lm}^{(\mathrm{sph})} = \sum_{i+j+k=l} t_{ijk,lm}^{(l)} \, \mathcal{I}_{ijk}^{(\mathrm{cart})},
\label{eq:cart2sph}
\end{equation}
where the coefficients $t_{ijk,lm}^{(l)}$ depend only on the angular momentum and can therefore be pretabulated or generated analytically~\cite{b8,b9}. Accordingly, each transformed index shrinks from $n_\mathrm{cart}(l) = (l+1)(l+2)/2$ Cartesian components to $n_\mathrm{sph}(l) = 2l+1$ pure spherical components.

The compression ratio grows rapidly with $l$: at $l=4$, each index shrinks from 15 to 9 components, yielding $15^4/9^4 = 7.7\times$ fewer contraction-bound intermediates for a four-index tensor. This exact shell transformation directly reduces Phase~2 node count, \texttt{atomicAdd} traffic to global memory, and downstream contraction work.

\medskip
\noindent\textbf{Stepwise Fusion into the Recurrence.}
A naive implementation would first materialize the full Cartesian intermediate tensor and only then apply Eq.~\eqref{eq:cart2sph}. FusionRCG instead fuses this standard shell transformation into Phase~2, applying it incrementally at HRR boundaries. The transformation state is tracked by a 4-bit mask, so each step retires a Cartesian index immediately after it has been consumed and replaces it with its smaller spherical counterpart. The result is a progressively shrinking intermediate tensor rather than a materialize-then-transform pipeline.

\begin{figure}[t]
\centering
\includegraphics[width=\columnwidth]{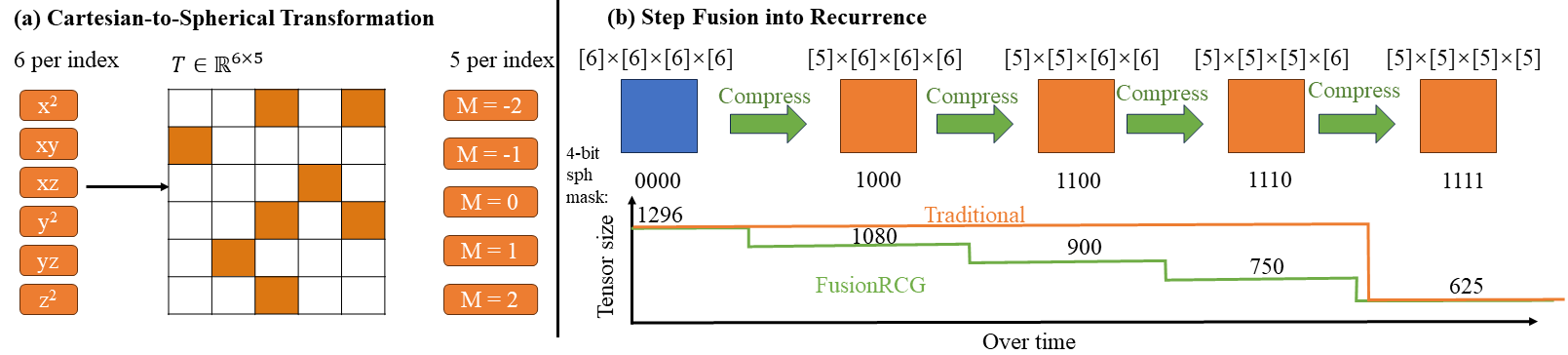}
\caption{Cartesian-to-spherical transformation and stepwise fusion. (a)~At $l=2$, the standard shell transformation maps 6 Cartesian Gaussian components to 5 pure spherical components per index via the fixed coefficient matrix $\mathbf{t}^{(l)}$. (b)~FusionRCG applies this transformation one index at a time under a 4-bit mask ($0000 \rightarrow 1000 \rightarrow 1100 \rightarrow 1110 \rightarrow 1111$), shrinking the intermediate tensor progressively instead of materializing the full Cartesian intermediate tensor first.}
\label{fig:sph}
\end{figure}

The correctness of this rearrangement follows from the fact that HRR operators are linear and act on disjoint index pairs. Let the contracted VRR output be the four-center Cartesian tensor
\begin{equation}
H_{\alpha\beta\gamma\delta}
:=
\big(
\chi^{A,\mathrm{cart}}_{\alpha}\,\chi^{B,\mathrm{cart}}_{\beta}
\,\big|\,
\chi^{C,\mathrm{cart}}_{\gamma}\,\chi^{D,\mathrm{cart}}_{\delta}
\big),
\end{equation}
where $\alpha,\beta,\gamma,\delta$ index Cartesian components on centers $A,B,C,D$. For a shell of angular momentum $l$, the corresponding pure spherical basis functions satisfy
\begin{equation}
\phi^{\mathrm{sph}}_{\mu}
:= \sum_{\alpha} T^{(l)}_{\mu\alpha}\,\chi^{\mathrm{cart}}_{\alpha}.
\end{equation}
The conventional scheme applies $\mathbf{T}^{(l_a)}{\otimes}\mathbf{T}^{(l_b)}{\otimes}\mathbf{T}^{(l_c)}{\otimes}\mathbf{T}^{(l_d)}$ after both HRR sweeps. FusionRCG interleaves these transformations with the bra--ket HRR sequence. For compact notation, the following derivation groups the two bra transformations and the two ket transformations; the implementation realizes the same algebra through the per-index mask shown in Figure~\ref{fig:sph}.

First, the bra-side HRR operator $\mathcal{R}_{ab}(\cdot;\mathbf{O}_{ab})$, with $\mathbf{O}_{ab}=\mathbf{A}-\mathbf{B}$, produces
\begin{equation}
G^{(1)}_{\alpha\beta\gamma\delta}
:= \big[\mathcal{R}_{ab}(\mathbf{H};\mathbf{O}_{ab})\big]_{\alpha\beta\gamma\delta}.
\end{equation}
The bra indices can then be transformed immediately:
\begin{equation}
G^{(2)}_{\mu\nu\gamma\delta}
:= \sum_{\alpha\beta}
T^{(l_a)}_{\mu\alpha}\,T^{(l_b)}_{\nu\beta}\,
G^{(1)}_{\alpha\beta\gamma\delta},
\label{eq:bra-sph}
\end{equation}
or compactly $\mathbf{G}^{(2)} = \bigl(\mathbf{T}^{(l_a)}{\otimes}\mathbf{T}^{(l_b)}\bigr)\mathbf{G}^{(1)}$. This step shrinks the bra half of the tensor from $n_\mathrm{cart}(l_a)n_\mathrm{cart}(l_b)$ to $n_\mathrm{sph}(l_a)n_\mathrm{sph}(l_b)$ components before the ket-side sweep begins.

The ket-side HRR operator $\mathcal{R}_{cd}(\cdot;\mathbf{O}_{cd})$, with $\mathbf{O}_{cd}=\mathbf{C}-\mathbf{D}$, then acts only on the ket indices:
\begin{equation}
G^{(3)}_{\mu\nu\gamma'\delta'}
:= \big[\mathcal{R}_{cd}(\mathbf{G}^{(2)};\mathbf{O}_{cd})\big]_{\mu\nu\gamma'\delta'}.
\end{equation}
Because $\mathcal{R}_{cd}$ is linear and touches only $(\gamma,\delta)$, while $\mathbf{T}^{(l_a)}{\otimes}\mathbf{T}^{(l_b)}$ acts only on $(\alpha,\beta)$, the two operations commute:
\begin{equation}
\mathcal{R}_{cd}\bigl(\mathbf{T}^{(l_a)}{\otimes}\mathbf{T}^{(l_b)}\bigr)
=
\bigl(\mathbf{T}^{(l_a)}{\otimes}\mathbf{T}^{(l_b)}\bigr)\mathcal{R}_{cd}.
\label{eq:sph-hrr-commute}
\end{equation}
Advancing the bra transformation ahead of the ket HRR therefore does not change the mathematical result. Finally, the ket indices are transformed:
\begin{equation}
G^{(4)}_{\mu\nu\lambda\sigma}
:= \sum_{\gamma'\delta'}
T^{(l_c)}_{\lambda\gamma'}\,T^{(l_d)}_{\sigma\delta'}\,
G^{(3)}_{\mu\nu\gamma'\delta'}.
\end{equation}

Chaining these steps gives the fused operator emitted by FusionRCG,
\begin{equation}
\mathbf{G}^{\mathrm{final}}
=
\bigl(\mathbf{T}^{(l_c)}{\otimes}\mathbf{T}^{(l_d)}\bigr)\,
\mathcal{R}_{cd}\,
\bigl(\mathbf{T}^{(l_a)}{\otimes}\mathbf{T}^{(l_b)}\bigr)\,
\mathcal{R}_{ab}\,
\mathbf{H}.
\label{eq:stepwise-sph}
\end{equation}
Using Eq.~\eqref{eq:sph-hrr-commute}, this is algebraically identical to the conventional materialize-then-transform pipeline,
\begin{equation}
\bigl(\mathbf{T}^{(l_a)}{\otimes}\mathbf{T}^{(l_b)}{\otimes}\mathbf{T}^{(l_c)}{\otimes}\mathbf{T}^{(l_d)}\bigr)\,
\mathcal{R}_{cd}\,\mathcal{R}_{ab}\,\mathbf{H}.
\label{eq:materialize-sph}
\end{equation}
Stepwise spherical fusion is therefore an implementation rearrangement, not a numerical approximation: it preserves the HGP result while avoiding the full Cartesian footprint during Phase~2.

\subsection{Adaptive Multi-Tier Kernel Architecture}
\label{sec:tiered}

Mathematically and algorithmically, the first two components reduce the transient live set and the size of Phase~2 intermediates, but they do not remove the strong dependence of total footprint on angular momentum. Architecturally, the remaining problem is therefore one of memory placement: different quartets must be mapped to different levels of the GPU hierarchy according to their decomposed register demand.

While the liveness-aware orchestration (\S\ref{sec:orchestration}) and algebraic dimensionality reduction (\S\ref{sec:spherical}) presented above significantly reduce on-chip memory pressure, the residual register footprint still varies dramatically across angular-momentum quartets. A natural design choice would be to adopt a single buffered execution pipeline for all cases, since buffering through shared memory and global memory can accommodate arbitrarily large working sets. However, such a uniform strategy introduces non-negligible fixed overheads---global buffer allocation, a layout-conversion kernel, and two separate kernel launches with synchronization barriers. For low-to-moderate degree quartets where total register spill remains below ${\sim}$1\,KB, these fixed costs exceed the spill penalty itself, making a lightweight register-only path consistently faster (validated in \S\ref{sec:ablation}). Conversely, for high-degree quartets whose persistent Phase~2 accumulator bank alone surpasses the hardware register budget, no scheduling improvement can fit the data on chip, and buffered execution becomes structurally mandatory.

This asymmetry motivates our \emph{adaptive two-tier} kernel architecture. The tier decision is derived from the decomposed register pressure model in Eq.~\eqref{eq:rtotal}: the same register file is jointly occupied by the transient VRR live frontier and the persistent accumulator bank (Figure~\ref{fig:hgp-bottleneck}), so the feasible execution regime depends on their combined footprint relative to the 255-register architectural limit. Concretely, this yields two hardware regimes:

\medskip
\noindent\textbf{Regime~I ($\sum l_i \le 6$).}
The \emph{combined} footprint of transient VRR intermediates and persistent accumulators still fits in registers. $N_\mathrm{acc}$ reaches at most 225 (at $l=(2,2,1,1)$: $5 \times 5 \times 3 \times 3$); after DSA, $\mathrm{PeakLive}_\mathrm{VRR} \lesssim 50$. Shared register occupancy therefore remains within the hardware budget with at most negligible spill ($<$1\,KB), so a register-only fused kernel is emitted (\S\ref{sec:tier1}).

\medskip
\noindent\textbf{Regime~II ($\sum l_i \geq 7$).}
The persistent accumulator bank already dominates the shared footprint. At $(2,2,2,1)$: $5 \times 5 \times 5 \times 3 = 375$ doubles $= 750$ registers---$3\times$ the hardware limit \emph{before any transient VRR term is added}. In this regime, keeping accumulators on chip is structurally impossible, so a buffered two-kernel pipeline is emitted (\S\ref{sec:tier2}) while VRR intermediates are still kept register-resident as much as possible.

The tier boundaries are not tuned hyperparameters: they follow directly from Eq.~\eqref{eq:rtotal} and the 255-register limit.

\subsubsection{Tier~1: Register-Only Kernels ($\sum l_i \le 6$)}
\label{sec:tier1}

When the shared footprint fits in registers (Regime~I), FusionRCG generates a \emph{single fused kernel} that keeps all intermediates in the register file.
Phase~1 (VRR) executes inside four nested primitive loops using DSA-scheduled register variables.
Results accumulate into Phase~2 (HRR) input variables via \texttt{+=}, exploiting the well-known accumulate-then-transform restructuring~\cite{b2} that avoids $n_p^4$ redundant HRR evaluations.
Phase~2 executes once after the loops, and the output contraction ($J/K$ matrix update via \texttt{atomicAdd}) is fused inline---no intermediate writes to global memory at any point.
Algorithm~\ref{alg:tier1} illustrates the generated kernel structure.

This path has zero fixed overhead: no global buffer allocation, no transpose, no extra kernel launch.
For quartets where residual spill is below $\sim$1\,KB (e.g., 484\,B for $(1,1,1,1)$), this overhead-free path outperforms the two-kernel Tier~2 pipeline despite the minor spill.

\begin{algorithm}[t]
\caption{Tier~1 Register-Only ERI Kernel (Optimized Stepwise Spherical Fusion)}
\label{alg:tier1}
\begin{algorithmic}[1]
\STATE \textbf{Input:} Shell quartet $(a,b,c,d)$ with primitives $\{(\alpha_i,c_i)\}$, density $D$
\STATE $\mathbf{H}[0 \ldots N_{\mathrm{hrr}}-1] \gets 0$ \hfill $\triangleright$ contraction accumulators (registers)
\FOR{each primitive quartet $(\alpha_a,\alpha_b,\alpha_c,\alpha_d)$}
  \STATE $\zeta,\eta,\mathbf{P},\mathbf{Q},T \gets \textsc{GaussianPairs}(\alpha_a,\alpha_b,\alpha_c,\alpha_d)$
  \STATE $F_0,\ldots,F_m \gets \textsc{Boys}(T,\,m)$
  \STATE $\mathbf{V} \gets \textsc{VRR}_{\pi^*}(F_0\ldots F_m,\,\zeta,\eta,\mathbf{P},\mathbf{Q})$ \hfill $\triangleright$ DSA-ordered schedule $\pi^*$
  \STATE $\mathbf{H} \mathrel{+}= c_a c_b c_c c_d \cdot \mathbf{V}$
\ENDFOR
\STATE $\mathbf{G}_{\mathrm{cart}} \gets \textsc{HRR}_{\mathrm{bra}}(\mathbf{H},\,\mathbf{O}_{ab})$ \hfill $\triangleright$ stepwise spherical fusion \\
\quad $\mathbf{G}_{\mathrm{sph}} \gets (\mathbf{T}^{(l_a)} \otimes \mathbf{T}^{(l_b)}) \cdot \mathbf{G}_{\mathrm{cart}}$ \\
\quad $\mathbf{G}_{\mathrm{cart}} \gets \textsc{HRR}_{\mathrm{ket}}(\mathbf{G}_{\mathrm{sph}},\,\mathbf{O}_{cd})$ \\
\quad $\mathbf{G}_{\mathrm{sph}} \gets (\mathbf{T}^{(l_c)} \otimes \mathbf{T}^{(l_d)}) \cdot \mathbf{G}_{\mathrm{cart}}$
\STATE $J,K \mathrel{+}= \textsc{Scatter}(\mathbf{G}_{\mathrm{sph}},\, D)$
\end{algorithmic}
\end{algorithm}

\subsubsection{Tier~2: Buffered Two-Kernel Pipeline ($\sum l_i \geq 7$)}
\label{sec:tier2}

When $N_\mathrm{acc}$ alone exceeds the register budget (Regime~II), accumulator variables \emph{cannot} reside in registers regardless of VRR scheduling.
This structural constraint---not an implementation choice---necessitates a fundamentally different architecture with two cooperating kernels (Figure~\ref{fig:overview} (a)).

\medskip
\noindent\textbf{Kernel~1 (Evaluation).}
A warp-sized batch of 32 threads cooperatively evaluates their respective shell quartets.
VRR intermediates remain in registers (bounded by DSA); HRR accumulators are written to an interleaved global buffer layout, enabling coalesced memory access across the warp.
A two-pass lifetime allocator (forward dry-run for birth/death timestamps, reverse-time best-fit packing) enables slot reuse: for $\mathbf{l}=(2,2,2,2)$, this compresses 3,545 slots to 1,865---a $1.9\times$ reduction.


\medskip
\noindent\textbf{Kernel~2 (Contraction).}
Each thread block processes one task with 64 threads.
Both the density matrix tile $D$ and the intermediate buffer are loaded into shared memory ($\le 48$\,KB), delivering $\sim$19\,TB/s bandwidth---$10\times$ over HBM.
The contraction $J_{cd} \mathrel{+}= \sum_{ab} D_{ab} \cdot I[a,b,c,d]$ exploits 8-fold ERI permutation symmetry via scale factors.

\section{Implementation}
\label{sec:codegen}

FusionRCG is implemented as an offline code generator rather than a hand-written collection of CUDA kernels. This design is necessary because each angular-momentum quartet induces a different recurrence graph, liveness profile, tier decision, and lowering strategy; encoding these combinations manually would be both error-prone and difficult to maintain as the optimization rules evolve. For each parameter tuple $\mathbf{l} = (l_1, l_2, l_3, l_4)$, the generator therefore executes a nine-stage pipeline: (1)~select tier from Eq.~\eqref{eq:rtotal}; (2)~construct Phase~1 computation graph using the axis-priority rule; (3)~schedule with DSA and insert barriers for Tier~2; (4)~[Tier~2] dry-run lifetime analysis and buffer slot packing; (5)~build Phase~2 computation graph with stepwise Cartesian-to-spherical fusion; (6)~schedule Phase~2; (7)~generate the accumulate-then-transform loop structure; (8)~emit specialized CUDA kernel (\texttt{.cu}); (9)~compile to device object with \texttt{nvcc}.

\begin{figure}[t]
\centering
\includegraphics[width=\columnwidth]{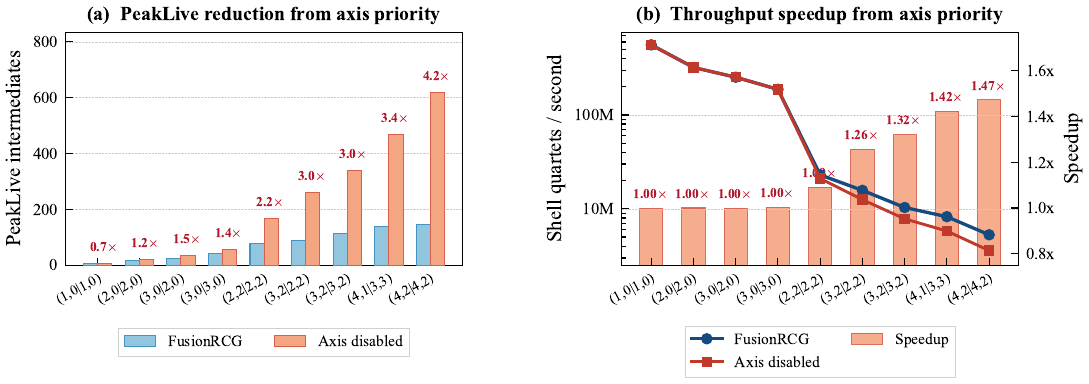}
\caption{Ablation of axis-priority graph construction across six quartets ($\sum l_i = 2$--$10$). (a)~PeakLive reduction factors. (b)~Throughput gain from axis priority.}
\label{fig:ablation-axis}
\end{figure}

\medskip
\noindent\textbf{Kernel specialization.}
For Tier~1, the final contraction is fused inline with zero intermediate global writes.
For Tier~2, the dedicated \texttt{\_jk\_transform} kernel tiles the density matrix~$D$ into shared memory and parallelizes the contraction across 64 threads per block, exploiting 8-fold ERI permutation symmetry via compile-time scale factors.

\medskip
\noindent\textbf{Build system.}
The parameter space has $\binom{l_\mathrm{max}+4}{4}$ unique angular-momentum configurations (120 for $l_\mathrm{max}=4$).
Including first-derivative variants and both tier architectures, the full library comprises 565 kernel files.
Generation is embarrassingly parallel: each $\mathbf{l}$-configuration is independent, so the pipeline is distributed across MPI ranks.
On a 32-core node, the complete library (generation + \texttt{nvcc} compilation) builds in several hours.

\section{Experimental Evaluation}
\label{sec:experiments}

\subsection{Experimental Setup}

\noindent\textbf{Platform.} All experiments run on NVIDIA A100-80GB GPUs (sm\_80, CUDA~12.4). Single-GPU evaluations use one A100; multi-GPU scaling tests use up to 64~GPUs across 8~nodes.

\medskip
\noindent\textbf{Baseline.}
We compare against GPU4PySCF~\cite{b3} (GPU-accelerated PySCF), the current state-of-the-art GPU quantum chemistry framework for Hartree--Fock and DFT calculations.
GPU4PySCF employs Rys-quadrature-based CUDA kernels and serves as the strongest publicly available GPU baseline for end-to-end SCF performance.
All reported timing results are averaged over three independent runs.

\subsection{Ablation Study}
\label{sec:ablation}

To validate each optimization component in isolation, we generate variant kernel libraries from the same code generator by toggling one component at a time: \emph{Common-way axis} (the conventional $l_a{\to}l_c$, $x{\to}y{\to}z$ ordering), \emph{No DSA} (a safe but non-optimized topological order), and \emph{No spherical} (spherical fusion disabled).
We also compare \emph{Tier~1-only} vs.\ \emph{Tier~2-only} execution to motivate adaptive routing.
Each variant is compiled as a complete library so that compiler reports and hardware counters reflect the true fused code path.
\subsubsection{Effect of Axis-Priority Graph Construction}
\label{sec:exp-axis}

Figure~\ref{fig:ablation-axis} isolates the axis-priority rule (Algorithm~\ref{alg:axis-priority}) by comparing the full pipeline against the conventional ordering ($l_a{\to}l_c$, $x{\to}y{\to}z$), labeled ``Axis disabled'' in the figure.
Both variants use the same DSA schedule; all differences arise solely from the graph topology choice.

The benefit scales consistently with angular momentum (Figure~\ref{fig:ablation-axis}).
At low $\sum l_i$ both topologies fit within the register budget and the throughput difference is marginal.
Once register pressure becomes significant ($\sum l_i \ge 7$), axis priority compresses PeakLive by up to $4.2\times$ and improves throughput by up to $+1.47\%$, confirming that the gain concentrates precisely where spill cost is highest.

\subsubsection{Effect of DSA Scheduling}

\begin{figure}[t]
\centering
\includegraphics[width=\columnwidth]{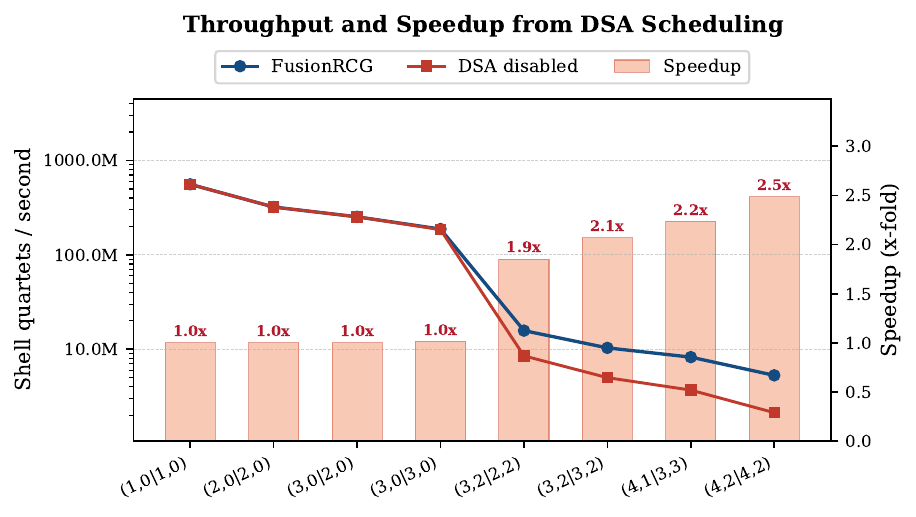}
\caption{Ablation of DSA scheduling across six quartets ($\sum l_i = 4$--$10$). Throughput and speedup from DSA. }
\label{fig:ablation-dsa}
\end{figure}

Figure~\ref{fig:ablation-dsa} isolates DSA by comparing the full pipeline against a variant using a safe but non-optimized topological order on the same graph structure.

For light quartets ($\sum l_i \le 6$), any reasonable topological order fits within the register file and DSA adds little (Figure~\ref{fig:ablation-dsa}).
Once spilling sets in ($\sum l_i \ge 8$), DSA delivers up to $2.3\times$ throughput gain and reduces dynamic local-memory traffic by up to $142\times$, confirming the causal chain: optimized evaluation order $\to$ fewer compiler spills $\to$ reduced off-chip traffic $\to$ higher throughput.

\subsubsection{Effect of Spherical Fusion}

\begin{figure}[t]
\centering
\includegraphics[width=\columnwidth]{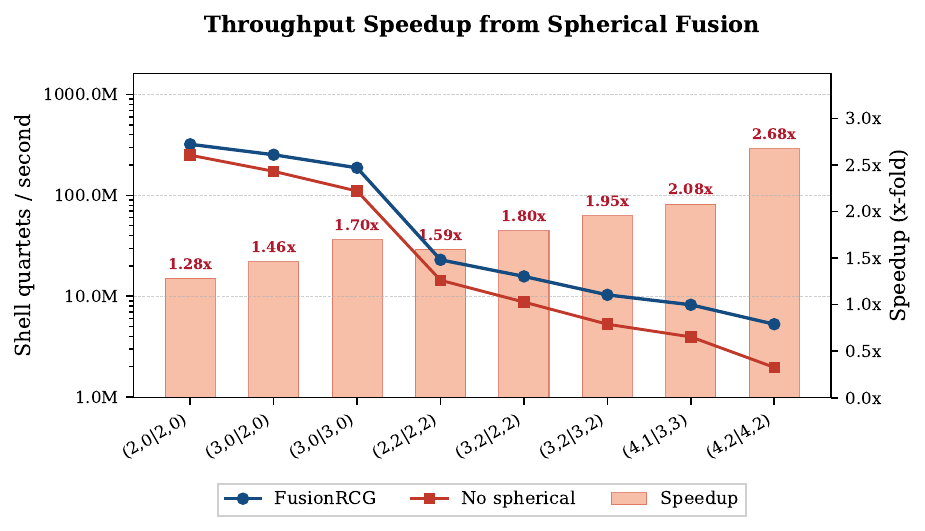}
\caption{Ablation of stepwise spherical fusion across six quartets ($\sum l_i = 4$--$10$). Throughput speedup over the no-spherical variant.}
\label{fig:ablation-spherical}
\end{figure}

Figure~\ref{fig:ablation-spherical} compares the full pipeline (with stepwise spherical fusion) against a variant operating entirely in the Cartesian basis (``No spherical'').

At low angular momentum the Cartesian-to-spherical size gap is small and the effect is negligible (Figure~\ref{fig:ablation-spherical}).
As $l$ increases, spherical fusion eliminates a growing fraction of floating-point operations---up to $12.81\%$ FLOPs reduction at the highest tested quartet---translating to a peak throughput speedup of $1.91\times$.
The stepwise strategy is key: each index is contracted and projected immediately rather than materializing the full Cartesian tensor, keeping the peak intermediate at the current-stage size and eliminating redundant arithmetic at the algebraic level.

\subsubsection{Effect of Adaptive Multi-Tier Routing}

\begin{figure}[t]
\centering
\includegraphics[width=\columnwidth]{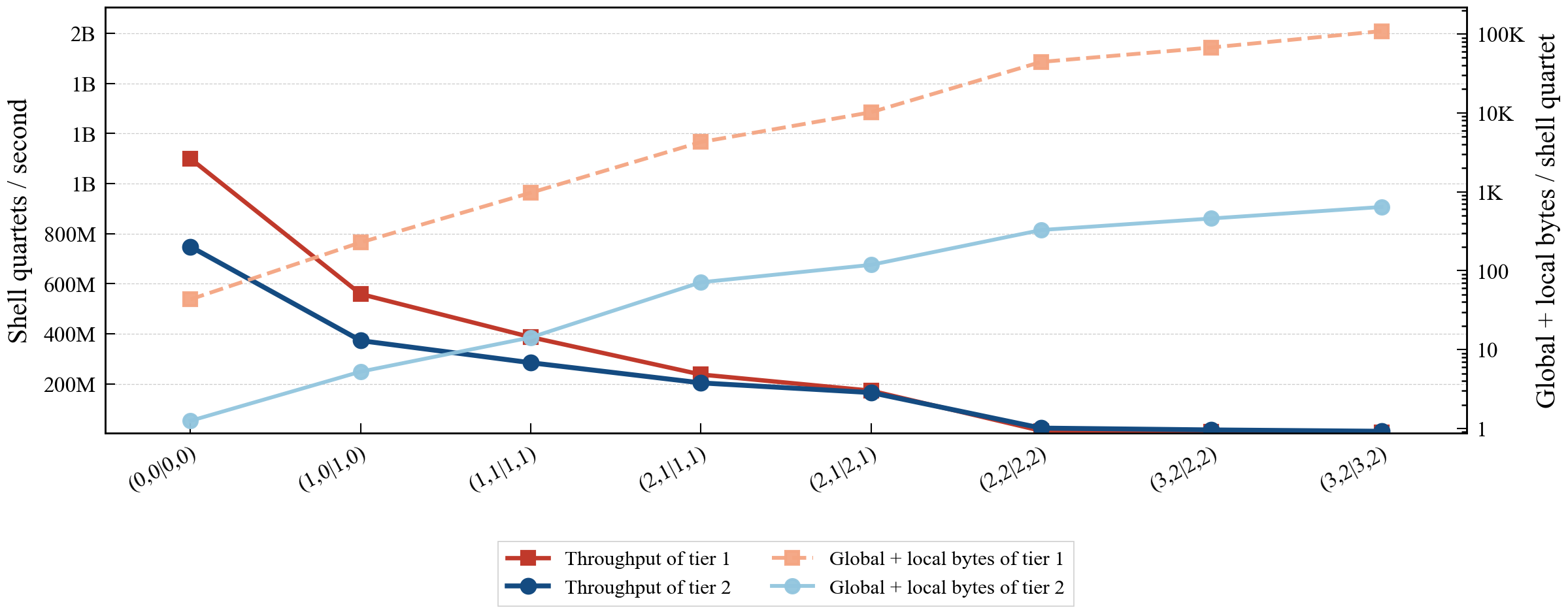}
\caption{Tier~1-only vs.\ Tier~2-only throughput over $\sum l_i = 0$--$16$. At low $\sum l_i$ Tier~1 wins due to zero pipeline overhead; at high $\sum l_i$ Tier~2 wins by eliminating register spills. Tier~1 fails to compile beyond $\sum l_i = 12$.}
\label{fig:ablation-tiered}
\end{figure}

Figure~\ref{fig:ablation-tiered} compares Tier~1-only (register-only for all quartets) against Tier~2-only (buffered pipeline for all quartets) across $\sum l_i = 0$--$16$.

At low angular momentum ($\sum l_i \le 6$), Tier~1 outperforms Tier~2 because the entire working set fits in registers, making Tier~2's additional pipeline stages (buffer allocation, layout conversion, dual kernel launch) unnecessary overhead.
As $\sum l_i$ grows, Tier~1's register working set overflows the 255-register limit, causing massive spilling; Tier~2 avoids this because its buffered two-kernel design keeps the per-thread memory footprint far lower, absorbing the overflow through shared and global memory rather than register spills.
Beyond $\sum l_i = 12$, Tier~1 fails to compile entirely.
This crossover motivates FusionRCG's adaptive routing: Tier~1 for $\sum l_i \le 6$ and Tier~2 for $\sum l_i \ge 7$, a boundary that follows directly from Eq.~\eqref{eq:rtotal}.

Taken together, the four ablations confirm that axis-priority graph construction and DSA scheduling jointly control register pressure and spill traffic, spherical fusion reduces both arithmetic and memory footprint at high~$l$, and adaptive tiering routes each quartet to the execution regime where it performs best. These components are complementary: their combined effect in the full FusionRCG pipeline exceeds any single optimization in isolation.

\subsection{End-to-End SCF Performance}
\label{sec:e2e}

We compare the time per SCF iteration against GPU4PySCF on a single A100.
Since GPU4PySCF tightly integrates ERI evaluation within its SCF solver, isolated kernel extraction is infeasible; we therefore compare at the application level, where both systems execute the identical SCF calculation on each molecular system.

\begin{figure}[t]
\centering
\includegraphics[width=\columnwidth]{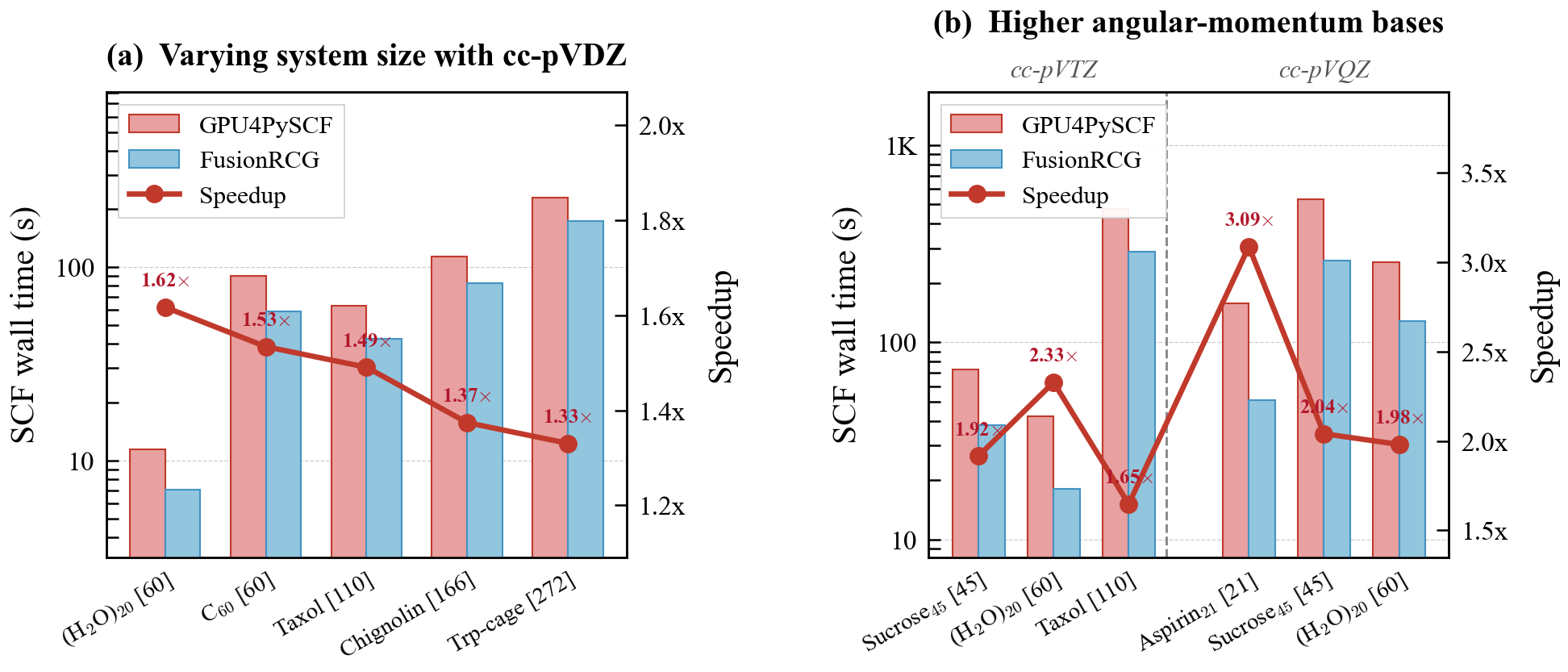}
\caption{Time per SCF iteration: GPU4PySCF vs.\ FusionRCG on a single A100 GPU. (a)~Varying system size with cc-pVDZ ($l_\mathrm{max}=2$): five molecular systems from 60 to 272 atoms. (b)~Higher angular-momentum basis sets: cc-pVTZ ($l_\mathrm{max}=3$) and cc-pVQZ ($l_\mathrm{max}=4$) on three systems each. The red line (right axis) shows the speedup of FusionRCG over GPU4PySCF.}
\label{fig:scf-comparison}
\end{figure}

\medskip
\noindent\textbf{Varying system size at fixed basis (cc-pVDZ).}
Figure~\ref{fig:scf-comparison}(a) evaluates five molecular systems of increasing size---from the water cluster (H$_2$O)$_{20}$ to the mini-protein Trp-cage (272 atoms)---with the cc-pVDZ basis set ($l_\mathrm{max}=2$).
FusionRCG consistently outperforms GPU4PySCF, achieving $1.33$--$1.62\times$ speedup across the entire range.
The advantage is largest on small systems and gradually narrows as system size grows, because cc-pVDZ involves primarily low angular-momentum shells and non-ERI SCF components (e.g., diagonalization, density fitting) constitute a growing fraction of each iteration.

\medskip
\noindent\textbf{Higher angular-momentum basis sets.}
Figure~\ref{fig:scf-comparison}(b) demonstrates FusionRCG's increasing advantage as the basis set incorporates higher angular-momentum functions.
With cc-pVTZ ($l_\mathrm{max}=3$), speedups range from $1.65\times$ to $2.33\times$.
Moving to cc-pVQZ ($l_\mathrm{max}=4$), where $f$- and $g$-type shell quartets become prevalent, FusionRCG achieves its peak advantage of $3.09\times$ on the drug cluster Aspirin$_{21}$, with all tested systems exceeding $1.98\times$.
This trend directly reflects the scaling of FusionRCG's optimizations: liveness-aware orchestration and spherical fusion provide progressively larger compression as angular momentum increases, while adaptive tiering becomes structurally essential for the $g$-function quartets that dominate cc-pVQZ workloads.

\begin{figure}[t]
\centering
\includegraphics[width=\columnwidth]{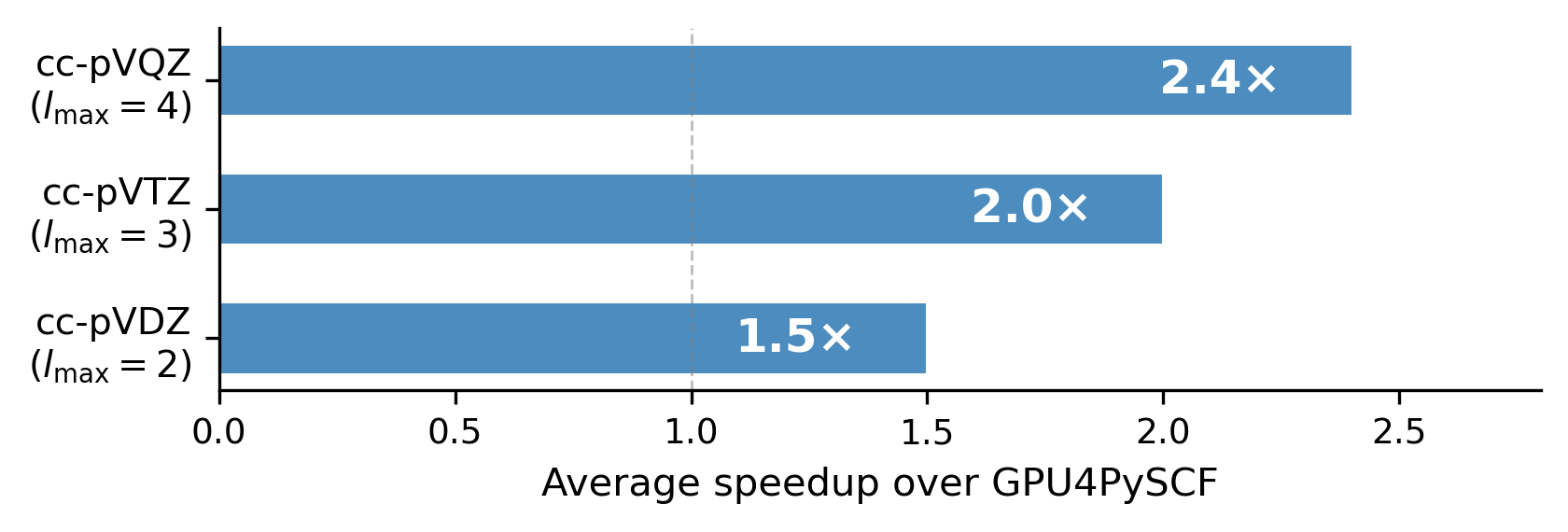}
\caption{Average SCF speedup of FusionRCG over GPU4PySCF across basis sets with progressively higher angular momentum. The advantage scales from $1.5\times$ (cc-pVDZ, $l_\mathrm{max}=2$) to $2.0\times$ (cc-pVTZ, $l_\mathrm{max}=3$) to $2.4\times$ (cc-pVQZ, $l_\mathrm{max}=4$).}
\label{fig:avg-speedup}
\end{figure}

\medskip
\noindent\textbf{Scaling trend across basis hierarchy.}
Figure~\ref{fig:avg-speedup} summarizes the aggregate trend: the average speedup over GPU4PySCF increases monotonically from $1.5\times$ at cc-pVDZ to $2.0\times$ at cc-pVTZ and $2.4\times$ at cc-pVQZ.
This consistent scaling validates our central thesis---that co-optimizing graph structure, intermediate representation, and memory mapping yields compounding benefits as the computational complexity of the recurrence grows.
The $2.4\times$ average improvement at cc-pVQZ, where high-$l$ integrals dominate the SCF cost, demonstrates that FusionRCG effectively rescues these workloads from the memory-bound regime.

\begin{figure}[t]
\centering
\includegraphics[width=\columnwidth]{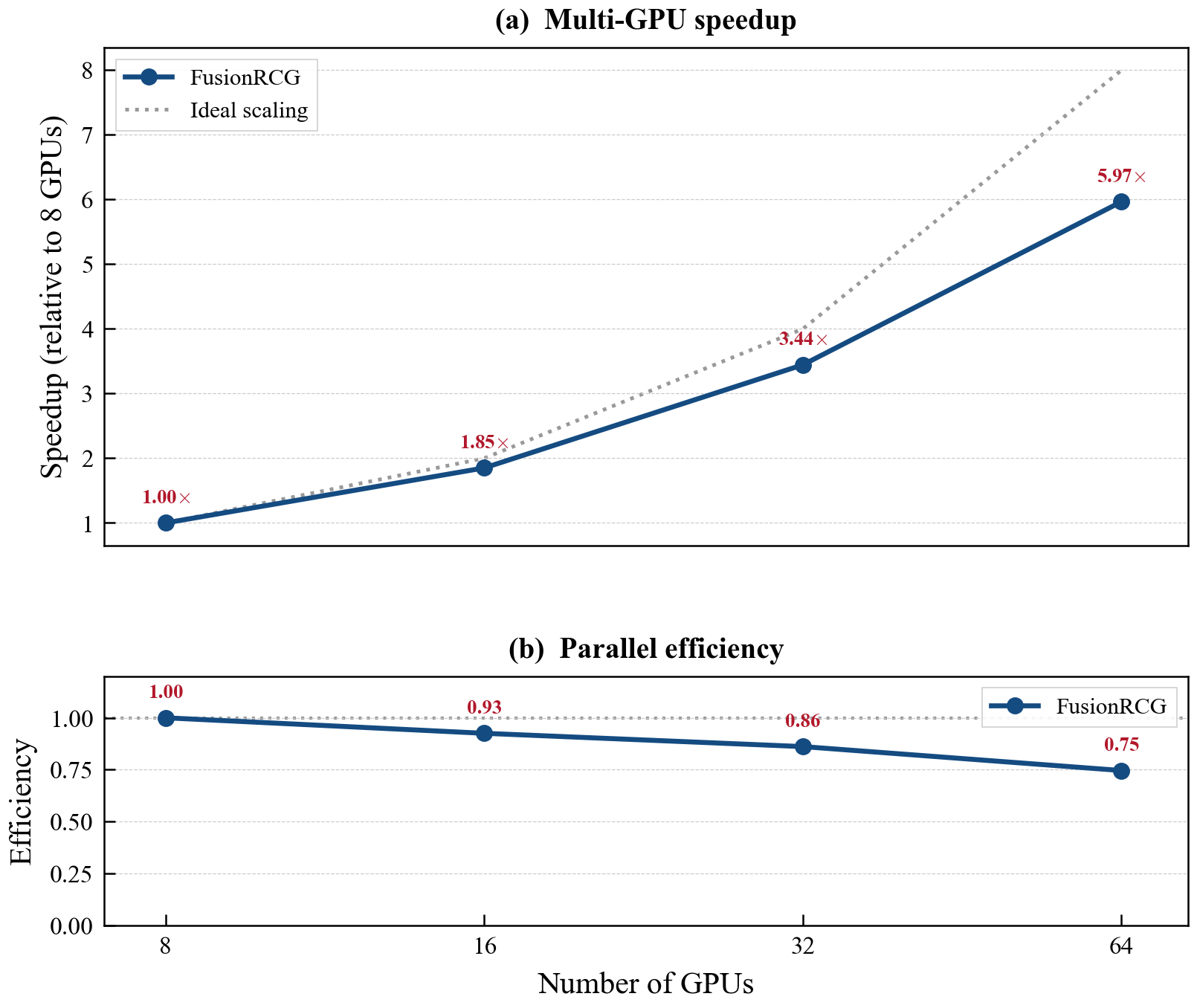}
\caption{Multi-GPU strong scaling on Ubiquitin (PDB:~1UBQ, 1{,}231 atoms)/cc-pVTZ. (a)~Speedup relative to 8~GPUs; the dashed line indicates ideal linear scaling. (b)~Parallel efficiency, defined as $\mathrm{speedup} / (\text{GPU count} / 8)$.}
\label{fig:scalability}
\end{figure}

To demonstrate that kernel-level register optimizations translate to system-level HPC scalability, we evaluate FusionRCG's strong scaling on a multi-node A100 cluster for Ubiquitin (PDB:~1UBQ~\cite{b36}, 1{,}231 atoms, 76 residues) with the cc-pVTZ basis set.
As shown in Figure~\ref{fig:scalability}, FusionRCG scales to $5.97\times$ speedup at 64~GPUs (8 nodes) relative to the 8-GPU baseline, maintaining $75\%$ parallel efficiency despite inter-node communication overhead and load imbalance among shell-quartet batches.
The sub-linear scaling is primarily attributable to the final $J/K$-matrix reduction across nodes and to workload heterogeneity among shell-quartet classes, which causes faster GPUs to idle while waiting for the slowest batch to complete.
This demonstrates that per-kernel register optimizations compose well with distributed-memory parallelism.

\section{Related Work}
\label{sec:related}

\noindent\textbf{GPU-accelerated quantum chemistry.}
TeraChem~\cite{b4} pioneered GPU ERI evaluation. While early implementations relied heavily on hand-tuned kernels for low angular momenta, subsequent developments have incorporated metaprogramming to support higher $l$; nevertheless, managing the exponential growth of register pressure remains a fundamental bottleneck.
GPU4PySCF~\cite{b3} provides an efficient CUDA-accelerated backend for PySCF and serves as our primary end-to-end baseline (\S\ref{sec:e2e}). Although it generates kernels for higher angular momenta, it relies heavily on local memory and experiences significant register spills, leading to degraded performance for complex basis sets.
LibintX~\cite{bLibintX} ports the widely used recursive ERI evaluation to GPUs, representing the most mature publicly available GPU-native ERI library for molecular integrals.
Mako~\cite{b12} reformulates ERIs as matrix multiplications to leverage Tensor Cores via CUTLASS, demonstrating impressive speedups. Though its closed-source nature precludes direct kernel-level comparison, it represents an orthogonal algorithmic approach to our thread-local recurrence method.
\emph{In contrast}, FusionRCG addresses the fundamental register-limit bottleneck of the recursive per-thread model by dynamically co-optimizing the algebraic recurrence paths and instruction scheduling for arbitrary $l$, enabling automated, high-performance kernel generation without manual tuning.

\subsection{Multi-GPU Scalability}
\label{sec:multigpu}

\medskip
\noindent\textbf{Graph scheduling and register pressure management.}
General-purpose minimum-register instruction scheduling is inherently NP-hard~\cite{b1}. Classical compiler heuristics~\cite{b15,b28,b16,b30} and GPU-specific optimizations, such as occupancy-aware instruction reordering~\cite{b29,b17} or kernel fission~\cite{b18}, operate on a \emph{fixed} dependence graph or scheduling region. Similar assumptions appear in dependency-rich GPU kernels such as sparse triangular solve, where an analysis phase first builds a graph from the matrix sparsity pattern and then executes that graph level by level~\cite{b31}. These methods improve execution order, synchronization, or resource usage on a prescribed DAG, but they do not change the graph itself.

HGP differs in two essential respects. First, the same target ERI admits multiple legal recurrence paths, so the computation graph is not a fixed input to scheduling: different axis choices induce genuinely different graph topologies and therefore different live ranges. Second, the dominant resource bottleneck is not merely dependency satisfaction on a fixed graph, but the overlap between a transient VRR live frontier and a persistent HRR accumulator bank under a hard per-thread register budget. General schedulers lack the domain information needed to exploit this algebraic degree of freedom. FusionRCG therefore couples graph construction with register-aware scheduling, reducing pressure \emph{before} the generated kernel reaches NVCC rather than only reordering a predetermined DAG.

\medskip
\noindent\textbf{Recursive and irregular GPU algorithms.}
A separate line of work studies recursive or highly irregular GPU workloads such as tree traversals, divide-and-conquer algorithms, and nested parallelism. For recursive tree traversals, prior work focuses on mitigating control and memory divergence across traversals by regrouping threads or by using hybrid CPU--GPU scheduling~\cite{b32}. For divide-and-conquer algorithms, GPU implementations such as QuickHull reformulate recursively generated subproblems as segmented data-parallel operations over contiguous array ranges~\cite{b33}. More generally, nested-parallel and dynamic-parallel approaches use child-kernel launches and workload consolidation to cope with runtime-generated work, while identifying kernel-launch overhead and GPU underutilization as the main difficulty~\cite{b34,b35}.

These techniques do not transfer directly to HGP. For a given shell quartet, the full HGP recurrence can be constructed statically at code-generation time; the core challenge is therefore not runtime task discovery, warp regrouping, or recursive kernel-launch granularity. Instead, HGP suffers from a compile-time, per-thread live-state explosion inside a small but dense algebraic recurrence DAG. Methods designed for segmented subproblem management, dynamic load balancing, or nested kernel consolidation do not reduce the VRR live frontier, the Phase~2 accumulator bank, or their overlap in the register file. FusionRCG targets this gap by treating HGP as a domain-specific recursive computation graph whose topology, schedule, representation, and memory placement must be co-optimized jointly.

\medskip
\noindent\textbf{Generation of Domain-specific code.}
Frameworks like ATLAS/FFTW~\cite{b19,b20} (for linear algebra and FFTs), Halide~\cite{b21}, and TVM~\cite{b22} (for deep learning) have demonstrated the immense power of auto-tuning and domain-specific compilation. While these tools excel at optimizing loop nests, memory tiling, and tensor operations, they are not designed to handle the combinatorial explosion of intermediate variables inherent in deeply recursive quantum chemistry integrals. 
\emph{FusionRCG uniquely expands} the scope of domain-specific compilation by tightly coupling mathematical recurrence generation, algorithmic path finding, and hardware-aware register management into a unified ERI code generator.

\section{Conclusion}

We have presented FusionRCG, a framework that addresses the GPU memory wall for hierarchical recursive computation graphs through three co-designed optimizations.
Our key insight---that the recurrence graph's structure is a degree of freedom, not a fixed constraint---enables liveness-aware graph orchestration that reduces peak simultaneously live intermediates by up to $3.2\times$ and boosts throughput by up to $44.7\%$.
Combined with stepwise spherical fusion that cuts FLOPs by up to $12.81\%$ and delivers up to $1.91\times$ kernel speedup, and an adaptive two-tier kernel architecture derived from register-pressure decomposition, FusionRCG achieves zero spill for all angular-momentum classes up to $\sum l_i = 6$ and extends feasible GPU execution to the full $g$-function range ($\sum l_i = 16$).
In end-to-end SCF comparison with GPU4PySCF on NVIDIA A100, FusionRCG achieves up to $3.09\times$ speedup, with the average advantage scaling from $1.5\times$ at cc-pVDZ to $2.4\times$ at cc-pVQZ---confirming that the benefit grows with angular-momentum complexity.
Furthermore, FusionRCG scales to 64~GPUs with $75\%$ parallel efficiency, demonstrating that per-kernel register optimizations compose effectively with distributed-memory parallelism.

Future directions include extending the framework to derivative integrals required for geometry optimization and molecular dynamics, and exploring Tensor Core co-execution for hybrid recurrence--matmul workloads.

\section*{Acknowledgment}
This project was supported by the Zhongguancun Academy under the Internal Research Grant No.~C20250501.

\bibliographystyle{IEEEtran}
\bibliography{references}

@IEEEtranBSTCTL{BSTcontrol,
  CTLuse_url = "yes",
  CTLdash_repeated_names = "no"
}

@article{b1,
  author  = {Ravi Sethi},
  title   = {Complete Register Allocation Problems},
  journal = {SIAM Journal on Computing},
  year    = {1975},
  volume  = {4},
  number  = {3},
  pages   = {226--248},
  url     = {https://doi.org/10.1137/0204020}
}

@article{b2,
  author  = {Martin Head-Gordon and John A. Pople},
  title   = {A Method for Two-Electron {G}aussian Integral and Integral Derivative Evaluation Using Recurrence Relations},
  journal = {The Journal of Chemical Physics},
  year    = {1988},
  volume  = {89},
  number  = {9},
  pages   = {5777--5786},
  url     = {https://doi.org/10.1063/1.455553}
}

@article{b3,
  author  = {Qiming Sun and Tianyu Zhu and Nicholas S. Blunt and others},
  title   = {Introducing {GPU} Acceleration into the {Python}-Based Simulations of Chemistry Framework},
  journal = {The Journal of Physical Chemistry A},
  year    = {2025},
  volume  = {129},
  number  = {5},
  pages   = {1459--1468},
  url     = {https://doi.org/10.1021/acs.jpca.4c05876}
}

@article{b4,
  author  = {Igor S. Ufimtsev and Todd J. Mart\'{\i}nez},
  title   = {Quantum Chemistry on Graphical Processing Units. 1. Strategies for Two-Electron Integral Evaluation},
  journal = {Journal of Chemical Theory and Computation},
  year    = {2008},
  volume  = {4},
  number  = {2},
  pages   = {222--231},
  url     = {https://doi.org/10.1021/ct700268q}
}

@article{b5,
  author  = {Samuel Williams and Andrew Waterman and David Patterson},
  title   = {Roofline: An Insightful Visual Performance Model for Multicore Architectures},
  journal = {Communications of the ACM},
  year    = {2009},
  volume  = {52},
  number  = {4},
  pages   = {65--76},
  url     = {https://doi.org/10.1145/1498765.1498785}
}

@inproceedings{b7,
  author    = {Andrew W. Appel and Lal George},
  title     = {Optimal Spilling for {CISC} Machines with Few Registers},
  booktitle = {Proceedings of the ACM SIGPLAN 2001 Conference on Programming Language Design and Implementation},
  year      = {2001},
  pages     = {243--253},
  url       = {https://doi.org/10.1145/378795.378854}
}

@article{b8,
  author  = {Frank Neese},
  title   = {The {SHARK} Integral Generation and Digestion System},
  journal = {Journal of Computational Chemistry},
  year    = {2023},
  volume  = {44},
  number  = {3},
  pages   = {381--396},
  url     = {https://doi.org/10.1002/jcc.26942}
}

@article{b9,
  author  = {Henry B. Schlegel and Michael J. Frisch},
  title   = {Transformation between Cartesian and Pure Spherical Harmonic {G}aussians},
  journal = {International Journal of Quantum Chemistry},
  year    = {1995},
  volume  = {54},
  number  = {2},
  pages   = {83--87},
  url     = {https://doi.org/10.1002/qua.560540202}
}

@misc{b10,
  author = {Edward F. Valeev},
  title  = {{Libint}: Machine-Generated Library for Efficient Evaluation of Molecular Integrals over {G}aussians},
  year   = {2025},
  url    = {https://evaleev.github.io/libint/}
}

@inproceedings{b12,
  author    = {Haozhi Han and Kun Li and Fusong Ju and Qi Li and Hong An and Yifeng Chen and Yunquan Zhang and Ting Cao and Mao Yang},
  title     = {Matrix Is All You Need: Rearchitecting Quantum Chemistry to Scale on {AI} Accelerators},
  booktitle = {Proceedings of the International Conference for High Performance Computing, Networking, Storage and Analysis},
  year      = {2025},
  pages     = {2126--2142},
  url       = {https://doi.org/10.1145/3712285.3759829}
}

@article{b13,
  author  = {Ravi Sethi and Jeffrey D. Ullman},
  title   = {The Generation of Optimal Code for Arithmetic Expressions},
  journal = {Journal of the ACM},
  year    = {1970},
  volume  = {17},
  number  = {4},
  pages   = {715--728},
  url     = {https://doi.org/10.1145/321607.321620}
}

@article{b15,
  author  = {Roberto {Casta{\~n}eda Lozano} and Christian Schulte},
  title   = {Survey on Combinatorial Register Allocation and Instruction Scheduling},
  journal = {ACM Computing Surveys},
  year    = {2019},
  volume  = {52},
  number  = {3},
  pages   = {62:1--62:50},
  url     = {https://doi.org/10.1145/3340313}
}

@article{b16,
  author  = {Sanghyun Park and Alex Nicolau and Aviral Shrivastava and Yunheung Paek and Nikil Dutt and Eugene Earlie},
  title   = {Bypass Aware Instruction Scheduling for Register File Power Reduction},
  journal = {ACM SIGPLAN Notices},
  year    = {2006},
  volume  = {41},
  number  = {7},
  pages   = {173--181},
  url     = {https://doi.org/10.1145/1159974.1134675}
}

@misc{b17,
  author = {Vasily Volkov},
  title  = {Better Performance at Lower Occupancy},
  year   = {2010},
  note   = {{GPU} Technology Conference},
  url    = {https://dmacssite.github.io/materials/volkov10-GTC.pdf}
}

@inproceedings{b18,
  author    = {Alok Sethia and Scott A. Mahlke},
  title     = {Equalizer: Dynamic Tuning of {GPU} Resources for Efficient Execution},
  booktitle = {Proceedings of the 47th Annual IEEE/ACM International Symposium on Microarchitecture},
  year      = {2014},
  pages     = {647--658},
  url       = {https://doi.org/10.1109/MICRO.2014.16}
}

@inproceedings{b19,
  author    = {R. Clint Whaley and Jack J. Dongarra},
  title     = {Automatically Tuned Linear Algebra Software},
  booktitle = {Proceedings of the ACM/IEEE Conference on Supercomputing},
  year      = {1998},
  url       = {https://doi.org/10.1109/SC.1998.10004}
}

@inproceedings{b20,
  author    = {Matteo Frigo and Steven G. Johnson},
  title     = {{FFTW}: An Adaptive Software Architecture for the {FFT}},
  booktitle = {Proceedings of the 1998 IEEE International Conference on Acoustics, Speech and Signal Processing},
  year      = {1998},
  volume    = {3},
  pages     = {1381--1384},
  url       = {https://doi.org/10.1109/ICASSP.1998.681704}
}

@inproceedings{b21,
  author    = {Jonathan Ragan-Kelley and Connelly Barnes and Andrew Adams and Sylvain Paris and Fr{\'e}do Durand and Saman Amarasinghe},
  title     = {{Halide}: A Language and Compiler for Optimizing Parallelism, Locality, and Recomputation in Image Processing Pipelines},
  booktitle = {Proceedings of the 34th ACM SIGPLAN Conference on Programming Language Design and Implementation},
  year      = {2013},
  pages     = {519--530},
  url       = {https://doi.org/10.1145/2499370.2462176}
}

@inproceedings{b22,
  author    = {Tianqi Chen and Thierry Moreau and Ziheng Jiang and others},
  title     = {{TVM}: An Automated End-to-End Optimizing Compiler for Deep Learning},
  booktitle = {13th USENIX Symposium on Operating Systems Design and Implementation},
  year      = {2018},
  pages     = {578--594},
  url       = {https://www.usenix.org/conference/osdi18/presentation/chen}
}

@article{b23,
  author  = {Henryk Laqua and J{\"o}rg Kussmann and Christian Ochsenfeld},
  title   = {Accelerating Seminumerical Fock-Exchange Calculations Using Mixed Single- and Double-Precision Arithmetic},
  journal = {The Journal of Chemical Physics},
  year    = {2021},
  volume  = {154},
  number  = {21},
  pages   = {214116},
  url     = {https://doi.org/10.1063/5.0045084}
}

@article{b24,
  author  = {Alexander Asadchev and Venkata Allada and Jason Felder and Brett M. Bode and Mark S. Gordon and Theresa L. Windus},
  title   = {Uncontracted {Rys} Quadrature Implementation of up to {G} Functions on Graphical Processing Units},
  journal = {Journal of Chemical Theory and Computation},
  year    = {2010},
  volume  = {6},
  number  = {3},
  pages   = {696--704},
  url     = {https://doi.org/10.1021/ct9005079}
}

@article{b25,
  author  = {Edward Palethorpe and Giuseppe M. J. Barca},
  title   = {High-Performance, High-Angular-Momentum {J} Engine on Graphics Processing Units},
  journal = {Journal of Chemical Theory and Computation},
  year    = {2025},
  volume  = {21},
  number  = {19},
  pages   = {9388--9403},
  url     = {https://doi.org/10.1021/acs.jctc.5c00775}
}

@article{b26,
  author  = {Gergely Samu and Mih{\'a}ly K{\'a}llay},
  title   = {Efficient Evaluation of Three-Center Coulomb Integrals},
  journal = {The Journal of Chemical Physics},
  year    = {2017},
  volume  = {146},
  number  = {20},
  pages   = {204101},
  url     = {https://doi.org/10.1063/1.4983393}
}

@article{b27,
  author  = {Alexander Asadchev and Edward F. Valeev},
  title   = {Implementation of {McMurchie--Davidson} Algorithm for {G}aussian {AO} Integrals Suited for {SIMD} Processors},
  journal = {The Journal of Physical Chemistry A},
  year    = {2025},
  volume  = {129},
  number  = {42},
  pages   = {9788--9797},
  url     = {https://doi.org/10.1021/acs.jpca.5c04136}
}

@article{b28,
  author  = {Christoph W. Kessler},
  title   = {Scheduling Expression {DAG}s for Minimal Register Need},
  journal = {Computer Languages},
  year    = {1998},
  volume  = {24},
  number  = {1},
  pages   = {33--53},
  url     = {https://doi.org/10.1016/S0096-0551(98)00002-2}
}

@inproceedings{b29,
  author    = {Ghassan Shobaki and Austin Kerbow and Stanislav Mekhanoshin},
  title     = {Optimizing Occupancy and {ILP} on the {GPU} Using a Combinatorial Approach},
  booktitle = {Proceedings of the 18th ACM/IEEE International Symposium on Code Generation and Optimization},
  year      = {2020},
  pages     = {133--144},
  url       = {https://doi.org/10.1145/3368826.3377918}
}

@inproceedings{b30,
  author    = {Ghassan Shobaki and John Bassett and Matthew Heffernan and Austin Kerbow},
  title     = {Graph Transformations for Register-Pressure-Aware Instruction Scheduling},
  booktitle = {Proceedings of the 31st ACM SIGPLAN International Conference on Compiler Construction},
  year      = {2022},
  pages     = {41--53},
  url       = {https://doi.org/10.1145/3497776.3517771}
}

@techreport{b31,
  author      = {Maxim Naumov},
  title       = {Parallel Solution of Sparse Triangular Linear Systems in the Preconditioned Iterative Methods on the {GPU}},
  institution = {NVIDIA Corporation},
  number      = {NVR-2011-001},
  year        = {2011},
  url         = {https://research.nvidia.com/publication/2011-06_parallel-solution-sparse-triangular-linear-systems-preconditioned-iterative}
}

@inproceedings{b32,
  author    = {Weiguo Liu and Bertil Schmidt and Wen{-}mei W. Hwu},
  title     = {Hybrid {CPU}-{GPU} Scheduling and Execution of Tree Traversals},
  booktitle = {Proceedings of the 21st ACM SIGPLAN Symposium on Principles and Practice of Parallel Programming},
  year      = {2016},
  pages     = {330--340},
  url       = {https://doi.org/10.1145/2851141.2851174}
}

@article{b33,
  author  = {Gang Mei and Nan Xu and Linbo Xu},
  title   = {A Sample Implementation for Parallelizing Divide-and-Conquer Algorithms on the {GPU}},
  journal = {Heliyon},
  year    = {2018},
  volume  = {4},
  number  = {3},
  pages   = {e00512},
  url     = {https://doi.org/10.1016/j.heliyon.2018.e00512}
}

@inproceedings{b34,
  author    = {Wenbin Li and Ghassan Shobaki and Tarek El-Ghazawi},
  title     = {Nested Parallelism on {GPU}: Exploring Parallelization Templates for Irregular Loops and Recursive Computations},
  booktitle = {2015 44th International Conference on Parallel Processing},
  year      = {2015},
  pages     = {595--604},
  url       = {https://doi.org/10.1109/ICPP.2015.107}
}

@inproceedings{b35,
  author    = {Ghassan Shobaki and Wenbin Li and Tarek El-Ghazawi},
  title     = {Compiler-Assisted Workload Consolidation for Efficient Dynamic Parallelism on {GPU}},
  booktitle = {2016 IEEE International Parallel and Distributed Processing Symposium},
  year      = {2016},
  pages     = {534--543},
  url       = {https://doi.org/10.1109/IPDPS.2016.110}
}

@misc{bLibintX,
  author = {Alexander Asadchev and Edward F. Valeev},
  title  = {{LibintX}: High-Performance Library for Scalable Molecular Integral Evaluation},
  year   = {2023},
  url    = {https://github.com/ValeevGroup/LibintX}
}

@article{b36,
  author  = {Senadhi Vijay-Kumar and Charles E. Bugg and William J. Cook},
  title   = {Structure of Ubiquitin Refined at 1.8~{{\AA}} Resolution},
  journal = {Journal of Molecular Biology},
  year    = {1987},
  volume  = {194},
  number  = {3},
  pages   = {531--544},
  url     = {https://doi.org/10.1016/0022-2836(87)90679-6}
}

\end{document}